\documentclass[twocolumn,showpacs,preprintnumbers,amsmath,amssymb,superscriptaddress,nofootinbib]{revtex4-1}
\usepackage{amsmath,amssymb,bm,graphicx,epsfig,psfrag,ulem,color,natbib}

\usepackage[abc]{overpic}

\newcommand{\vv}{{\bf {v}}}
\newcommand{\rr}{{\bf {r}}}
\newcommand{\pp}{{\bf {p}}}

\newcommand{\eF}{{\mbox{$\epsilon_F$}}}

\newcommand{\pF}{{\mbox{$p_F$}}}
\newcommand{\vg}{{\mbox{$v_g$}}}

\newcommand{\bnabla}{{\mbox{\boldmath $\nabla$}}}
\newcommand{\kb}{{\mbox{$k_B$}}}

\begin{document}

\title{\bf Scattering length of Andreev reflection from quantized vortices in $^3$He-$B$}

\author{Y. A. Sergeev}
\email{yuri.sergeev@ncl.ac.uk}
\affiliation{
Joint Quantum Centre Durham-Newcastle, and
School of Mechanical and Systems Engineering,
Newcastle University,
Newcastle upon Tyne, NE1 7RU, UK
}
\author{C. F. Barenghi}
\affiliation{
Joint Quantum Centre Durham-Newcastle, and
School of Mathematics and Statistics, Newcastle University,
Newcastle upon Tyne, NE1 7RU, UK
}

\author{S.~N.~Fisher}
\thanks{Deceased 4 January 2015}
\affiliation{Department of Physics, Lancaster University, Lancaster, LA1 4YB, UK}

\author{V.~Tsepelin}
\affiliation{Department of Physics, Lancaster University, Lancaster, LA1 4YB, UK}

\author{N. Suramlishvili}
\affiliation{School of Mathematics, University of Bristol, Bristol, BS8 1TW, UK
}

\date {\today}

\begin{abstract}
Andreev reflection of thermal quasiparticles from quantized vortices is an important technique to visualize quantum turbulence
in low temperature $^3$He-$B$. We revisit a problem of Andreev reflection from the isolated, rectilinear vortex line. For quasiparticle excitations whose impact parameters, defined as distances of the closest approach to the vortex core, do not exceed some arbitrary value $b$, we
calculate exactly the reflected fraction of the total flux of excitations incident upon the vortex in
the direction orthogonal to the vortex line. We then define and calculate exactly, as a function of $b$, the scattering length, that is the scattering cross section per unit length of the vortex line. We also define and calculate the scattering lengths for the flux of energy carried by thermal excitations, and for the net energy flux resulting from a (small) temperature gradient, and analyze the dependence of these scattering lengths on temperature.
\end{abstract}

\maketitle

\section{Introduction}
\label{sec:intro}

In the zero temperature limit, a pure superfluid can often be modeled
as an inviscid fluid~\cite{Donnelly-book}. The origin of superfluidity
in $^4$He is the Bose-Einstein condensation of helium atoms into the
quantum ground state. In liquid $^3$He-$B$, which is a Fermi superfluid,
the condensate is formed by Cooper pairs of Fermi atoms, the collective
behavior of which is described by the coherent macroscopic wavefunction
$\Psi(\rr,t)=\vert \Psi (\rr,t)\vert e^{i\theta(\rr,t)}$, where
$\theta$ is the phase, $\rr$ position, and $t$ time.
The superfluid velocity $\vv$ is then proportional to $\bnabla\theta$,
hence superfluid motion is irrotational, $\bnabla\times\vv={\bf 0}$ everywhere except for
topological defects on which $\bnabla\times\vv$ is singular.
Such line defects are quantized vortex lines. Around each of them the
phase of the wavefunction changes by $2\pi$ hence giving rise to
irrotational circulating flow.
The circulation around each vortex line is quantized, with quantum of
circulation being $\kappa=h/(2 m_3)=\pi \hbar/m_3=0.662\times10^{-7}\,{\rm m^2/s}$, where $2m_3$ and $m_3$ are, respectively, the mass of a Cooper pair and a bare $^3$He atom.
In the zero temperature limit, where the presence
of the normal fluid can be ignored, each of the quantized vortices moves
with the local fluid velocity in the flow field generated by the collective
flow field of all other vortices~\cite{Donnelly-book,BDV,BDV-book}.
When helium is stirred, this results in the formation of a
vortex tangle; such a tangle, together with the flow field generated by it,
is known as quantum turbulence.

It is worth mentioning here more exotic kinds of quantized vortices that may exist in $^3$He but are not observed in other known superfluids. For example, in their review~\cite{Bunkov-review}, Bun'kov {\it et al.} described observations of the nonsingular vortex textures in the rotating A phase. In another very substantial review of quantized vortices in $^3$He \cite{Salomaa-review} (see also references therein) Salomaa and Volovik discussed unique vortex phenomena such as continuous coreless vortices with two flow quanta and vortices with a half-integer circulation quanta in the A phase, and the vortex-core phase transition which results in a spontaneous bifurcation of vorticity involving half-quantum vortices in the B phase.

At low temperatures unique properties of $^3$He-$B$ permit quantum turbulence to be noninvasively probed by a minuscule normal component. Experiments have demonstrated that the ambient normal component comprises a gas of ballistic excitations at temperatures below $0.3T_c$ (superfluid transition temperature $T_c$ is 0.9\,mK at zero pressure)~\cite{Corney}. While these ballistic thermal excitations do not affect the motion of quantized vortices, they are Andreev scattered (reflected) from the flow field of quantized vortices, see, e.g., Refs.~\cite{Fisher_review,PNAS} and references therein. The Andreev reflection of thermal excitations is used to develop experimental techniques for measurements of quantum turbulence in the low temperature $^3$He-$B$ (see, e.g., Refs.~\cite{Fisher2001,Hosio}).

Here we describe briefly the mechanism of Andreev reflection~\cite{Andreev} in the
Fermi superfluid. In a quiescent fluid, the dispersion curve, $E(\pp)$
for excitations with energy $E$ and momentum $\pp$, has a minimum at
the Fermi momentum, $\pF=8.25\times10^{-25}\,{\rm kg\,m/s}$ which
corresponds to the Cooper pair binding energy (superfluid energy gap)
$\Delta$. Thermal excitations with $p>\pF$ are called quasiparticles,
and those with $p<\pF$ are called quasiholes (here $p=\vert\pp\vert$).
Consider now a vicinity of the minimum of the dispersion curve.
On moving from one side of the minimum to the other, the
excitations' group velocity
$\vg=dE/dp$ reverses sign, that is, quasiparticles and
quasiholes whose momenta are similar travel in opposite directions.
Consider now a superfluid moving with velocity $\vv$. In the reference
frame of the moving superfluid, the dispersion curve will be tilted
by the
Galilean transformation~\cite{Fisher_review}
\begin{equation}
E(\pp)\to E(\pp)+\pp\cdot\vv\,,
\label{eq:Galilean}
\end{equation}
so that quasiparticles propagating into the region where the superfluid
velocity $\vv$ has the same direction as their momentum $\pp$
experience a potential barrier. Travelling at constant energy,
a quasiparticle with insufficient energy
will be Andreev reflected as a quasihole whose path will practically
coincide with that of the incident quasiparticle \cite{BSS1}.
For quasiparticles with higher energies the flow field will be
transparent. Quasiholes are either Andreev reflected or transmitted
in the opposite way. It is important to notice
that the momentum transfer from incident thermal excitations to the flow
is negligibly small, so that the Andreev reflection offers a practically
noninvasive tool for probing the quantized vorticity at low temperatures.

The Andreev reflection from vortex configurations can be characterized by its
scattering cross section. In general this cross section depends on the area of impact parameters of quasiparticle excitations incident upon the region of quantized vorticity. In this paper we will consider the Andreev reflection from an isolated, rectilinear vortex line: The simple geometry allows us to find asymptotic solutions analytically. The case of curved vortex lines must be solved numerically~\cite{tangle}. Furthermore, in our case where the incident excitations propagate in the direction orthogonal to the vortex filament, the considered problem becomes two dimensional and the Andreev reflection can be characterized by the scattering length, that is the scattering cross section per unit length of the vortex line. The scattering length will depend upon the size of the considered interval of impact parameters, the latter defined as the distances of the closest approach of a quasiparticle excitation to the vortex core. The {\it scale} of this scattering length (although not the scattering length itself) was estimated (see Refs.~\cite{Fisher_review,PNAS} and references therein) to be very large, of the order of a few microns at the lowest accessible temperature 0.1$T_c$ ($100\,\mu{\rm K}$ at zero pressure).
The fortunate consequence of this result is that, since the estimated radius of the vortex core -- the distance over which $\Psi$ changes from its bulk value to zero -- is of the order of $10^{-7}~\rm m$ (see, e.g., Ref.~\cite{Tsuneto}), our problem does not depend on the detailed nature of the vortex core.

There are two different experimental realizations~\cite{Fisher_review} of turbulence measurements using the Andreev scattering technique. One experimental setup, in which ambient quasiparticles are detected by a vibrating wire surrounded by turbulence, enables the experimentalist to measure the reflected `particle' (or `number') flux of excitations. The other setup makes use of the black body radiator which generates the beam of quasiparticles propagating through a region of quantized vorticity. In this realization, a detector measures the reflected (or transmitted) energy carried by thermal excitations. Note that in the case where the temperature is uniform, the density of the energy flux is balanced by an identical flux density of quasiparticles traveling in the opposite direction; the nonuniformity of temperature results in the net energy flux carrying by excitations traveling in opposite directions.

Below in Sec.~\ref{sec:numberflux}, for thermal excitations, incident upon the vortex, whose impact parameters (distances of the closest approach to the vortex core) do not exceed an arbitrary maximum value $b$, we calculate exactly the Andreev-reflected fraction of the total flux. We then define and calculate exactly, as a function of $b$, the scattering length of Andreev reflection. In Sec.~\ref{sec:energy} we define the scattering lengths for the flux of energy carried by thermal excitations, and for the
net energy flux resulting from a (small) temperature gradient, and analyze their dependence on $b$ and temperature. In Sec.~\ref{sec:conclusions} we summarize our results and draw the
conclusions.

\section{Andreev scattering from an isolated rectilinear vortex}
\label{sec:numberflux}

We revisit the problem of ballistic propagation of thermal quasiparticle excitations through the superflow field of the quantized vortex in $^3$He-$B$ at low temperatures, $T<0.3T_c$.
In polar coordinates $(r,\,\phi)$ in the plane orthogonal to the vortex filament the superfluid velocity field of an isolated, rectilinear, quantized vortex is
\begin{equation}
\vv=\frac{\kappa}{2\pi r}\,{\bf e}_\phi\,,
\label{eq:v}
\end{equation}
where ${\bf e}_\phi$ is the azimuthal unit vector.

Andreev reflection~\cite{Andreev} by an isolated quantized vortex, described in more detail in Refs.~\cite{Fisher_review,PNAS,BSS1,Enrico,Fisher1989}, is illustrated schematically in
 \begin{figure}[ht]
 \begin{center}
 \epsfig{file=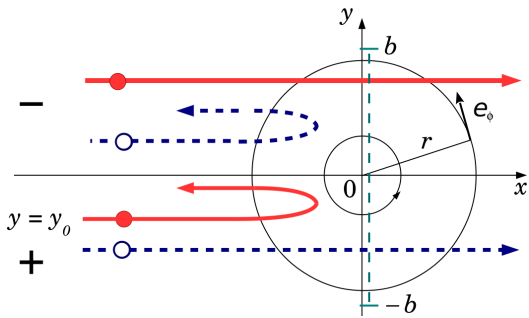,width=0.8\linewidth,clip=}
 \end{center}
 \caption{(Color online) Schematic illustration of the Andreev reflection by a
rectilinear quantized vortex; small solid (red) circles and thick red lines are
quasiparticles and their trajectories, small open (blue)
circles and thick dashed (blue) lines are quasiholes and their trajectories,
$y_0$ is the $y$ coordinate of an incident excitation. In regions ``$+$'' ($y<0$) and ``$-$'' ($y>0$), $\pp\cdot\vv>0$ and $\pp\cdot\vv<0$, respectively. The vertical dashed green line
shows the interval of the $y$ coordinates such that the impact parameters $\vert y_0\vert$ do not exceed $b$.}
 \label{fig:1}
 \end{figure}
Fig.~\ref{fig:1}. The superflow~(\ref{eq:v}) generates a local energy
barrier to thermal excitations. Consider an excitation incident
on the vortex from the left, defined in the figure as the $x$ direction. Since the Andreev-reflected excitation
practically retraces the path of the incident quasiparticle~\cite{BSS1}
(note that in Fig.~\ref{fig:1} the turning regions of quasiparticle
trajectories are not shown in scale), the trajectories of
quasiparticles with momentum $p>\pF$ and energy $E>\eF$,
where $\eF=2.27\times10^{-23}\,{\rm J}$ is the Fermi energy, and of
quasiholes, with $p<\pF$ and $E<\eF$, can be identified by their
`initial' $y$ coordinates, $y_0$. Low energy quasiparticles and
quasiholes are Andreev reflected, respectively, on the opposite sides
of the vortex.

The one-dimensional `particle' (or `number') flux density of excitations (that is, a number of quasiparticle excitations per unit area per unit time) incident on the flow field of the vortex is \cite{Fisher_review,PNAS,Fisher1989}
\begin{equation}
\langle n\vg\rangle_i=\int\limits_\Delta^\infty\vg(E)G(E)f(E)\,dE=G(\pF)\kb Te^{-\Delta/k_BT}\,,
\label{eq:flux}
\end{equation}
where $\Delta$ is the superfluid energy gap, $\vg(E)$ is the excitation group velocity, $G(E)$ is the density of states, and $f(E)$ is the Fermi distribution which, in the low temperature limit, reduces to the Boltzmann distribution, $f(E)=\exp(-E/\kb T)$; the integral in Eq.~(\ref{eq:flux}) is calculated noting that $\vg(E)G(E)=G(\pF)$, where $G(\pF)$ is the density of momentum states at the Fermi surface.

In what follows, rather than $y_0$ it is more convenient to regard as an
impact parameter the distance of the closest approach of
quasiparticle excitation to the vortex core, $b=\vert y_0\vert>0$.
Consider now quasiparticle excitations with $p>\pF$.
The density of the flux transmitted through the ``lower'' half-plane of the
flow field (region ``$+$'' in Fig.~\ref{fig:1}), $y<0$
in which $\pp\cdot\vv>0$, can be calculated by means of the integral
in Eq.~(\ref{eq:flux}) with the lower limit replaced,
in accordance with Eq.~(\ref{eq:Galilean}),
by $\Delta+\max\vert\pp\cdot\vv\vert$, where
$\max\vert\pp\cdot\vv\vert$ is determined along the
quasiparticle's rectilinear trajectory $y=y_0$.
Since $p\approx\pF$, making use of Eq.~(\ref{eq:v})
and $\kappa=2\pi\hbar/(2 m_3)$ we find $\max\vert\pp\cdot\vv\vert=\pF\hbar/2m_3b$
(note that this value corresponds to the maximum tilt of the dispersion
curve in the reference frame of the moving superfluid).
For considered excitations with $p>\pF$, the ``upper'' half-plane
(region ``$-$'' in Fig.~\ref{fig:1}), $y<0$ in which
$\pp\cdot\vv<0$, is transparent. Therefore, for the density of transmitted
flux we obtain~\cite{Fisher_review,PNAS}
\begin{eqnarray}
\langle n\vg\rangle^{+}_t &=& \int\limits_{\Delta+p_F\hbar/2m_3b}^\infty  \vg(E)G(E)f(E)\,dE \nonumber \\
&=& G(\pF)\kb T e^{-\Delta/k_BT} e^{-b_0/b}\,, \nonumber \\
\langle n\vg\rangle^{-}_t &=& G(\pF)\kb Te^{-\Delta/k_BT}\,,
\label{eq:fluxt}
\end{eqnarray}
where the superscripts $+$ and $-$ indicate corresponding regions of the flow field, and
\begin{equation}
b_0=\frac{\pF\hbar}{2m_3\kb T}\,.
\label{eq:b0}
\end{equation}

The quasihole excitations ($p<\pF$) are transmitted in the opposite way: for these excitations the region $y<0$ is transparent, and the flux density transmitted through the flow field of the vortex is determined by Eqs.~(\ref{eq:fluxt}) in which $\langle n\vg\rangle^{+}_t$ should be replaced by $\langle n\vg\rangle^{-}_t$ and vice versa.

For quasiparticles, the probability of transmission (that is, the transmitted fraction of the incident flux density), $f_t^\pm=\langle n\vg\rangle^{\pm}_t/\langle n\vg\rangle_i$ follows immediately from Eq.~(\ref{eq:fluxt}) as
\begin{equation}
f_t^+(b)=\exp(-b_0/b)\,, \quad f_t^-(b)=1\,.
\label{eq:ft}
\end{equation}
As seen from Eq.~(\ref{eq:ft}), the quantity $b_0$, defined by Eq.~(\ref{eq:b0}), sets the {\it scale} of  Andreev scattering length (but not the scattering length itself -- see below). The value of $b_0$ is about $6.3\,\mu{\rm m}$ at $T=100\,\mu{\rm K}$ and zero pressure and exceeds by several orders of magnitude the superfluid coherence length, $\xi_0\approx50\,{\rm nm}$. Such a large value of $b_0$ length results from the slow, $1/r$ decrease of the superfluid velocity with the distance from the vortex core whose thickness is of the order of $\xi_0$. In the following analysis the thickness of the vortex core is neglected and the vortex filament is regarded as the infinitesimally thin vortex line.

To define formally and then determine the Andreev scattering length, we start with the general, three-dimensional case and consider the one-dimensional flux of thermal excitations propagating in a certain direction and incident upon the region of quantized vorticity. Impact parameters of thermal excitations can be defined as (`initial') coordinates of quasiparticles on a plane orthogonal to the direction of propagation. For excitations whose impact parameters are within some surface area, say $S$, the cross section of Andreev reflection can be defined as
\begin{equation}
\sigma=S\langle q\rangle_r/\langle q\rangle_i\,,
\label{eq:sigma}
\end{equation}
where
\begin{equation}
\langle q\rangle_i=S\langle n\vg\rangle_i\,, \quad \langle q\rangle_r=\int\limits_S\langle n\vg\rangle_r\,dS
\label{eq:q3D}
\end{equation}
are the total incident and reflected fluxes of thermal excitations through the surface area $S$, and $\langle n\vg\rangle_r=\langle n\vg\rangle_i-\langle n\vg\rangle_t$ is the reflected flux density. Clearly, the scattering cross-section depends on the area $S$ of impact parameters.

For rectilinear vortex lines, in the case where the incident one-dimensional flux of excitations propagates in the direction, say $x$, which is orthogonal to vortex filaments, the problem of Andreev scattering becomes two dimensional, and instead of scattering cross section we can introduce the scattering length $K$ defined as the scattering cross section per unit length of the vortex line. In the system of coordinates $(x,\,y)$, consider excitations whose ``initial'' $y$ coordinates are within some interval $c\leq y_0\leq d$. For these excitations, similar to Eqs.~(\ref{eq:sigma}) and (\ref{eq:q3D}), the scattering length of Andreev reflection is defined as
\begin{equation}
K=(d-c)\langle q\rangle_r/\langle q\rangle_i\,,
\label{eq:K}
\end{equation}
where now
\begin{equation}
\langle q\rangle_i=(d-c)\langle n\vg\rangle_i\,, \quad \langle q\rangle_r=\int\limits_c^d\langle n\vg\rangle_r\,dy
\label{eq:q2D}
\end{equation}
are the total incident and reflected fluxes of thermal excitations through the interval $[c,\,d]$ of the $y$ axis. In general, the scattering length depends on $c$ and $d$.

To calculate the scattering length for an isolated, rectilinear, quantized vortex we consider the scattering of thermal excitations whose distances of the closest approach to the vortex core (that is, their impact parameters) do not exceed some arbitrary value $b$. Then, in definition~(\ref{eq:K}), $c=-b$ and $d=b$ (see Fig.~\ref{fig:1}, where the interval of $y$ coordinates such that the impact parameters do not exceed $b$ is shown by the vertical dashed green bar), so that
\begin{equation}
\langle q\rangle_i=2b\langle n\vg\rangle_i=2bG(\pF)\kb Te^{-\Delta/k_BT}\,.
\label{eq:intfluxi}
\end{equation}
Making use of Eqs.~(\ref{eq:fluxt}) we find that $\langle n\vg\rangle^{-}_r=0$ and
\begin{equation}
\langle n\vg\rangle^{+}_r=G(\pF)\kb Te^{-\Delta/k_BT}\left(1-e^{-b_0/b}\right)\,,
\label{eq:nvgr+}
\end{equation}
so that $\langle q\rangle_r=\int_0^b\langle n\vg\rangle_r^+\,db$, and for the Andreev-reflected fraction, $F=\langle q\rangle_r/\langle q\rangle_i$ of the total flux (that is the flux density integrated in the considered interval of impact parameters) and for the scattering length $K$ of Andreev reflection, defined by Eq.~(\ref{eq:K}), we obtain
\begin{equation}
F(b)=\frac{1}{2}\left(1-\frac{1}{b}\int\limits_0^b e^{-b_0/b}\,db\right)\,, \quad K(b)=2bF(b)\,.
\label{eq:qr}
\end{equation}
Introducing the nondimensional impact parameter $z=b/b_0$, the Andreev-reflected fraction of the total flux can be represented in the universal (that is, free of nondimensional parameters) form:
\begin{equation}
F(z)=\frac{1}{2z}\int\limits_0^z\left(1-e^{-1/z}\right)\,dz\,.
\label{eq:Fzdef}
\end{equation}
The integral in Eq.~(\ref{eq:Fzdef}) can be expressed through the integral exponential function~\cite{Abramowitz}
\begin{equation}
E_1(x)=\int\limits_x^\infty\frac{e^{-x}}{x}\,dx
\label{eq:E1}
\end{equation}
in the form
\begin{equation}
F(z)=\frac{1}{2z}\left[z\left(1-e^{-1/z}\right)+E_1\left(\frac{1}{z}\right)\right]\,.
\label{eq:Fz}
\end{equation}
\begin{figure*}[ht]
\centering
\begin{tabular}{cc}
\epsfig{file=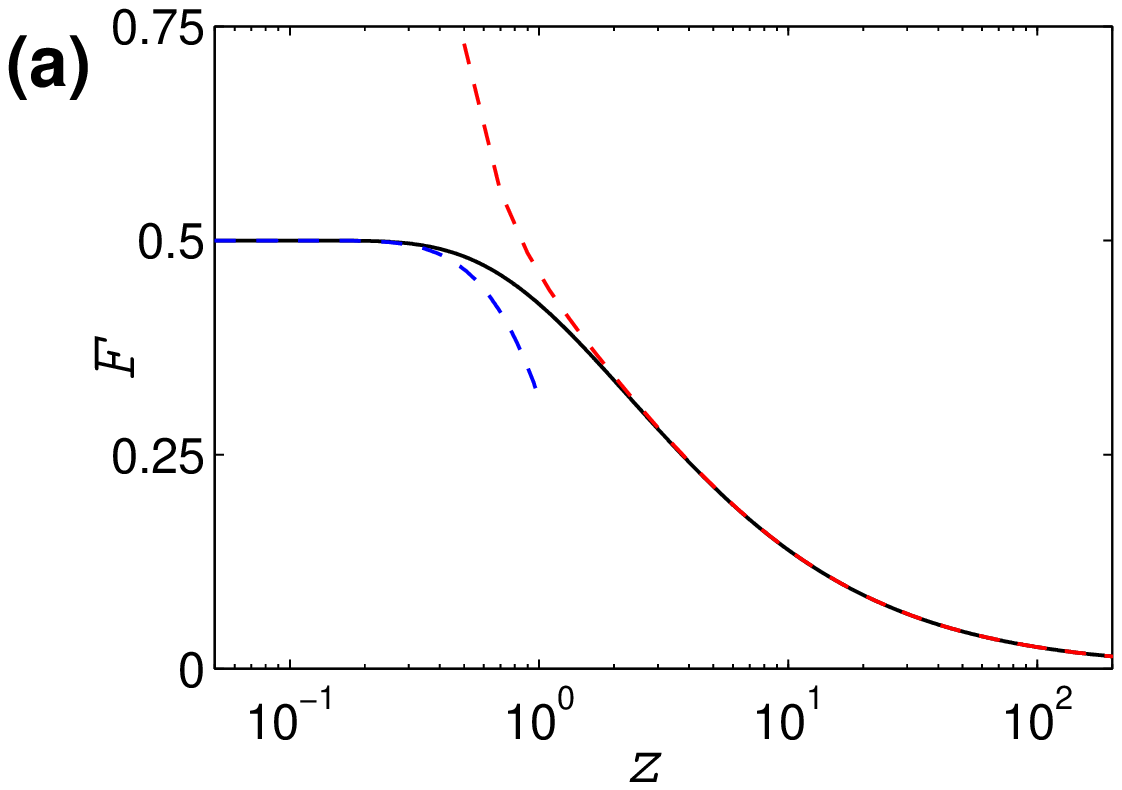,width=0.4\linewidth,clip=} &
\epsfig{file=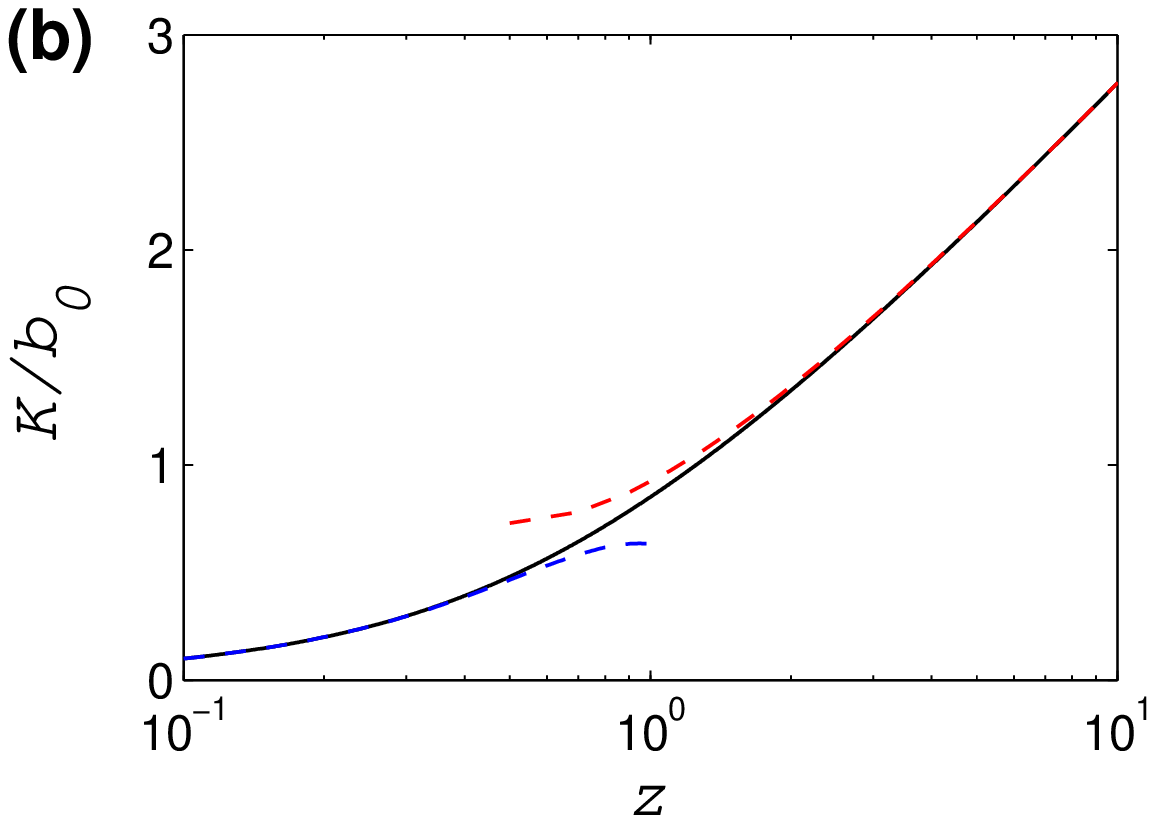,width=0.4\linewidth,clip=}
\end{tabular}
\caption{(Color online) (a) The reflected fraction of the total flux for impact parameters such that they do not exceed an arbitrary value $b$ and (b) the scattering length in units of the scale $b_0$ as functions of the nondimensional impact parameter $z=b/b_0$. Dashed (red and blue) lines show asymptotics~(\ref{eq:Fz-largez}) and (\ref{eq:Fz-smallz}) for large and small values of the impact parameter, respectively.}
\label{fig:2}
\end{figure*}
In units of $b_0$, the scattering length is also a universal function of the nondimensional impact parameter, $z=b/b_0$ alone, that is
\begin{equation}
\tilde{K}(z)=\frac{K}{b_0}=2zF(z)=z\left(1-e^{-1/z}\right)+E_1\left(\frac{1}{z}\right)\,.
\label{eq:Ktilde}
\end{equation}
Useful asymptotics~\cite{Abramowitz}, corresponding to large and small values of the impact parameter, are, respectively:
\begin{equation}
z=\frac{b}{b_0}\gg1: \quad \tilde{K}(z)=\ln z+1-\gamma+\frac{1}{2z}+\dots\,,
\label{eq:Fz-largez}
\end{equation}
where $\gamma=0.5772156649\dots$ is the Euler-Mascheroni constant, and
\begin{equation}
z=\frac{b}{b_0}\ll1: \quad \tilde{K}(z)=z\left(1-ze^{-1/z}\right)+\dots\,.
\label{eq:Fz-smallz}
\end{equation}
The reflected fraction $F$ of the total flux and the scattering length in units of the scale $b_0$ are shown in Fig.~\ref{fig:2} as functions of the nondimensional
impact parameter $z=b/b_0$. [For a wider range of $z$ the scattering length will also be shown in Fig.~\ref{fig:3}(b) of the next section.] A rather slow $\sim z^{-1}\ln z$ decrease of the reflected fraction $F=\tilde{K}/(2z)$ and the corresponding slow logarithmic increase of the scattering length with $z$ for large values of the impact parameter are due to the fact that a significant fraction of thermal excitations is still Andreev reflected even at large distances from the vortex core.

Note that for $T=100\,\mu{\rm K}$ and zero pressure the scattering length $K$ is about $0.8b_0\approx5\,\mu{\rm m}$ for $b=b_0$, and about $25\,\mu{\rm m}$ for $b=100b_0$. These very large values of scattering length should have a pronounced effect on the Andreev reflection from vortex configurations and vortex tangles, in particular on screening effects which were briefly discussed in our recent paper~\cite{PNAS}. However, we leave a more detailed discussion of these effects for future publications.

\section{Scattering length of the energy flux}
\label{sec:energy}
In Sec.~\ref{sec:numberflux}, for the particle (`number') flux of excitations incident on quantized vortex, we calculated exactly the reflected fraction of the total flux of excitations whose impact parameters do not exceed an arbitrary value $b$ and defined and calculated the corresponding scattering length. Of interest for interpretation of Andreev reflection experiments and numerical simulations would also be the scattering length defined for the flux of energy (rather than the number flux) carried by incident excitations. Similar to Eq.~(\ref{eq:flux}), the incident energy flux density is calculated as
\begin{eqnarray}
\langle n\vg E\rangle_i&=&\int\limits_\Delta^\infty\vg(E)G(E)f(E,\,T)E\,dE \nonumber \\
&=&G(\pF)\kb T(\Delta+\kb T)e^{-\Delta/k_BT}\,.
\label{eq:eflux}
\end{eqnarray}
The average energy of excitation in this flux is $\langle E\rangle=\Delta+\kb T$. For a uniform temperature this flux density is balanced by an identical flux density of quasiparticles traveling in the opposite direction.

Together with the energy flux density defined by Eq.~(\ref{eq:eflux}) we will also consider the density of the net energy flux, for excitations traveling in opposite directions, resulting from a small temperature difference $\delta T$ between two sides of the vortex flow field~\cite{BSS1}:
\begin{eqnarray}
\delta\langle n\vg E\rangle_i=\delta T\int\limits_\Delta^\infty\vg(E)G(E)\frac{\partial f(E,\,T)}{\partial T}E\,dE \nonumber \\
=\delta T\frac{G(\pF)}{\kb T}\left[(\Delta+\kb T)^2+(\kb T)^2\right]e^{-\Delta/k_BT}\,.
\label{eq:netflux}
\end{eqnarray}

We consider first the transmission and reflection of the energy flux defined by Eq.~(\ref{eq:eflux}). For the density of energy flux transmitted through the vortex we obtain
\begin{eqnarray}
&&\langle n\vg E\rangle^{+}_t = \int\limits_{\Delta+p_F\hbar/2m_3b}^\infty  \vg(E)G(E)f(E,\,T)E\,dE \nonumber \\
&&=G(\pF)\kb T(\Delta+\kb T)\left(1+\beta\frac{b_0}{b}\right) e^{-\Delta/k_BT} e^{-b_0/b}\,, \nonumber \\
&&\langle n\vg E\rangle^{-}_t = \langle n\vg E\rangle_i\,,
\label{eq:efluxt}
\end{eqnarray}
where
\begin{equation}
\beta=\frac{\kb T}{\Delta+\kb T}\,.
\label{eq:beta}
\end{equation}
For the reflected energy flux density we have $\langle n\vg E\rangle_r^+=\langle n\vg E\rangle_i-\langle n\vg E\rangle_t^+$. The scattering length is now defined as $K_E=2b\langle q_E\rangle_r/\langle q_E\rangle_i$, where $\langle q_E\rangle_i$ and $\langle q_E\rangle_r$ are the total incident and reflected energy fluxes, respectively, through the interval of impact parameters such that the distances of the closest approach of quasiparticles to the vortex core are not larger than the maximum value $b$. Repeating the steps leading, in Sec.~\ref{sec:numberflux}, from Eq.~(\ref{eq:flux}) to Eq.~(\ref{eq:Ktilde}) for the reflected fraction of the total energy flux $F_E=\langle q_E\rangle_r/\langle q_E\rangle_i$ and the corresponding scattering length, we obtain
\begin{equation}
F_E=\frac{1}{2z}\int\limits_0^z\left[1-\left(1+\frac{\beta}{z}\right)e^{-1/z}\right]\,dz\,, \quad K_E=2bF_E\,.
\label{eq:FEzdef}
\end{equation}
Making use of elementary transformations, the integral $\int_0^z\beta z^{-1}\exp(-1/z)\,dz$ can be expressed through the integral exponential function $E_1$ defined by Eq.~(\ref{eq:E1}), yielding the nondimensional scattering length $\tilde{K}_E=K_E/b_0$ in the form
\begin{equation}
\tilde{K}_E(z)=z\left(1-e^{-1/z}\right)+(1-\beta)E_1\left(\frac{1}{z}\right)\,.
\label{eq:FEz}
\end{equation}
The reflected fraction of the total energy flux is, obviously, $F_E(z)=\tilde{K}_E/(2z)$.

Proceeding in a similar way, for the net energy flux~(\ref{eq:netflux}) resulting from a small temperature gradient, we obtain for the transmitted flux density:
\begin{eqnarray}
\delta\langle n\vg E\rangle^{+}_t &=& \delta T\int\limits_{\Delta+p_F\hbar/2m_3b}^\infty  \vg(E)G(E)\frac{\partial f(E,\,T)}{\partial T}E\,dE \nonumber \\
&=&\delta T\frac{G(\pF)}{\kb T}\left[(\Delta+\kb T)^2+(\kb T)^2\right] \nonumber \\
&\times&\left(1+\beta_T\frac{b_0}{b}\right)^2 e^{-\Delta/k_BT} e^{-b_0/b}\,, \nonumber \\
\delta\langle n\vg E\rangle^{-}_t &=& \delta\langle n\vg E\rangle_i\,,
\label{eq:efnetluxt}
\end{eqnarray}
where
\begin{equation}
\beta_T=\frac{\kb T(\Delta+\kb T)}{(\Delta+\kb T)^2+(\kb T)^2}\,.
\label{eq:betaT}
\end{equation}
For the reflected flux density we have $\delta\langle n\vg E\rangle_r^+=\delta\langle n\vg E\rangle_i-\delta\langle n\vg E\rangle_t^+$. Similar to Eq.~(\ref{eq:FEzdef}), for the reflected fraction $F_T=\delta\langle q_T\rangle_r/\delta\langle q_T\rangle_i$ of the total net energy flux $\delta\langle q_T\rangle_i=2b\delta\langle n\vg E\rangle_i$ (here $\delta\langle q_T\rangle_r$ is the total reflected flux for quasiparticles with impact parameters which do not exceed $b$), we obtain the reflected fraction of the total incident flux as a function of the nondimensional impact parameter $z=b/b_0$:
\begin{equation}
F_T(z)=\frac{1}{2z}\int\limits_0^z\left[1-\left(1+\frac{\beta_T}{z}\right)^2e^{-1/z}\right]\,dz\,.
\label{eq:FTz}
\end{equation}
Calculating the integral in Eq.~(\ref{eq:FTz}), for the scattering length in units of $b_0$, and the reflected fraction of the net energy flux resulting from a temperature gradient, we obtain
\begin{eqnarray}
\tilde{K}_T&=&\frac{K_T}{b_0} \nonumber \\
&=&z\left(1-e^{-1/z}\right)+(1-2\beta_T)E_1\left(\frac{1}{z}\right)-\beta_T^2e^{-1/z}\,, \nonumber \\ 
F_T(z)&=&\frac{\tilde{K}_T}{2z}\,.
\label{eq:KTz}
\end{eqnarray}

If $T\to0$ then $\beta=\beta_T=0$ and the scattering lengths $K_E$ and $K_T$ of the energy fluxes reduce to the scattering length $K$ of the number flux, see Eq.~(\ref{eq:Ktilde}). This is consistent with the results of our earlier paper~\cite{SBBS4} in which we found from numerical simulations that for temperatures $T\leq0.15T_c$ the `thermal' and particle cross sections of Andreev scattering from vortex rings are almost indistinguishable.

The calculations performed in this paper are valid only for the ballistic regime of propagation of thermal excitations, that is within the range of temperatures $0\leq T\leq0.3T_c$, see, e.g., Ref.~\cite{Corney}. The temperature dependencies of the scattering lengths $K_E$ and $K_T$ are determined by the $1/T$ behavior of the scale $b_0$ [see Eq.~(\ref{eq:b0})] as well as by the behavior with temperature of the parameters $\beta$ and $\beta_T$ defined by Eqs.~(\ref{eq:beta}) and (\ref{eq:betaT}), respectively. Note that the temperature dependence of the superfluid energy gap can be neglected: The solution of the gap equation~\cite{Culler,DeGennes} shows that within the considered temperature interval $0\leq T\leq0.3T_c$ the gap $\Delta(T)$ changes by only about 1\%. It can be safely assumed that in the ballistic regime the superfluid energy gap is temperature independent, and $\Delta=1.76\kb T_c$~\cite{Bardeen,Tsuneto}.

\begin{figure*}[ht]
\centering
\begin{tabular}{cc}
\epsfig{file=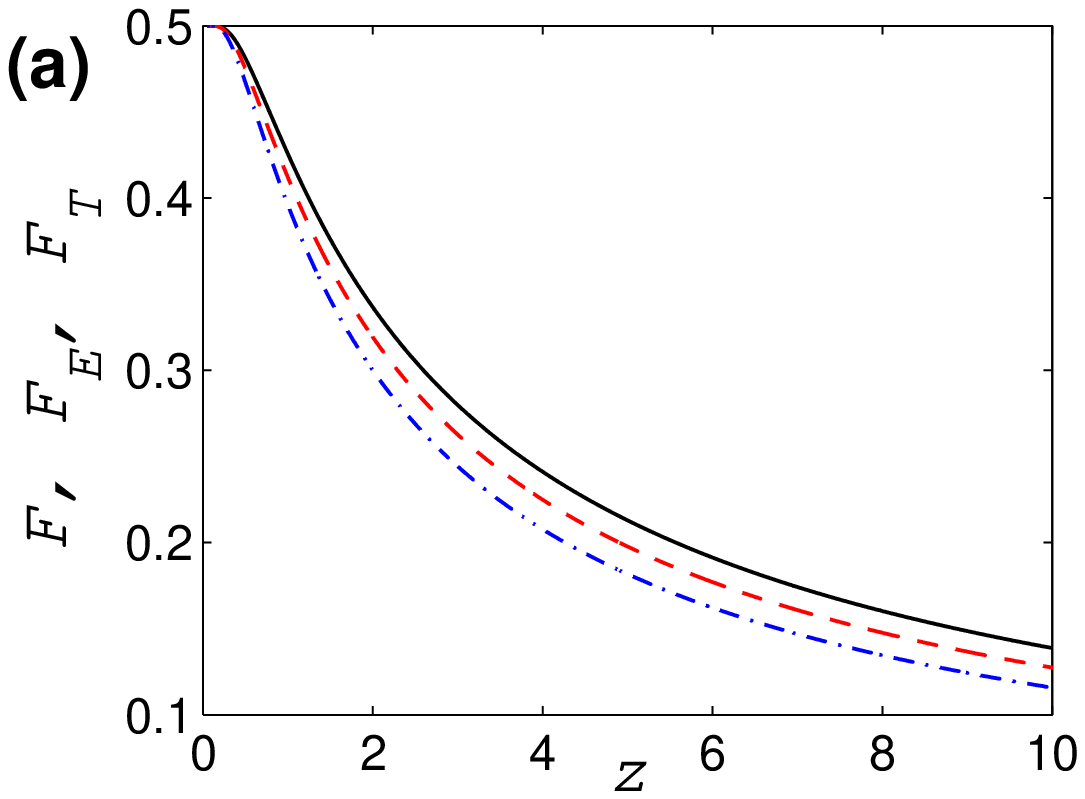,width=0.37\linewidth,clip=} &
\epsfig{file=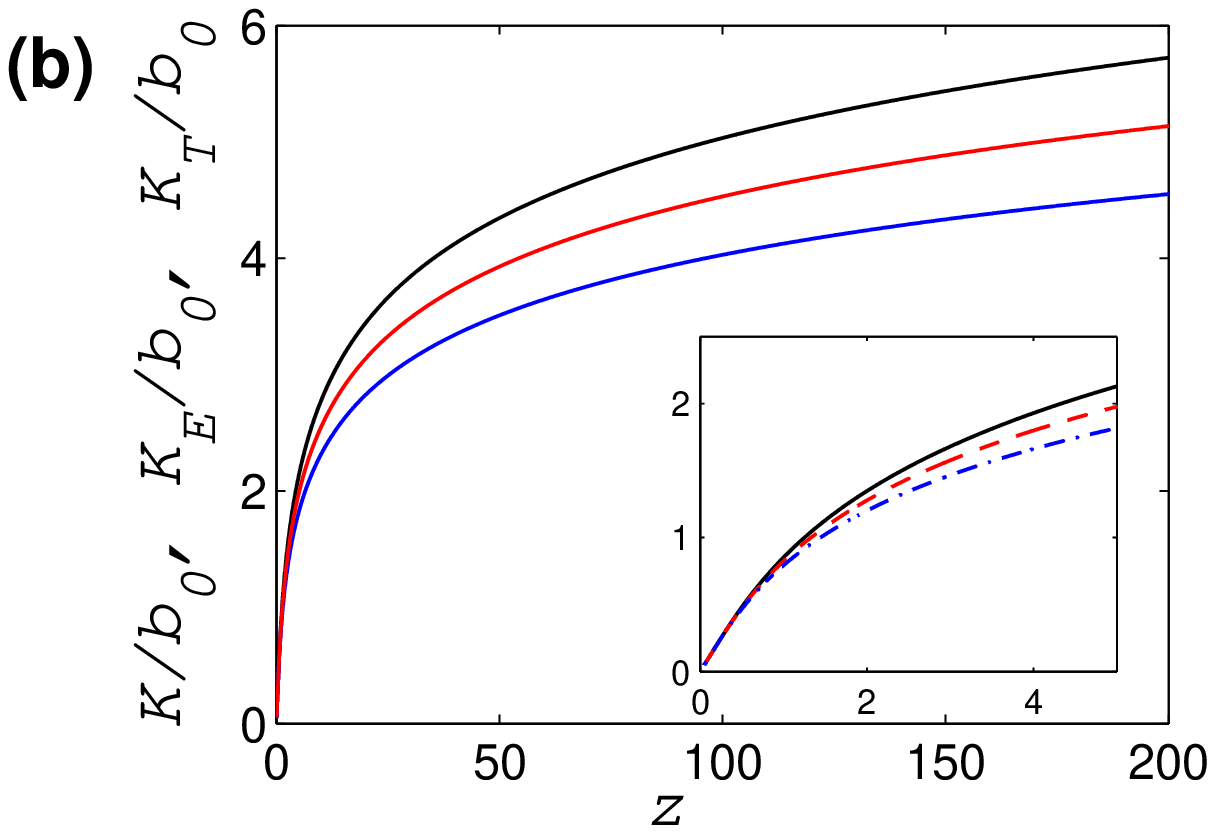,width=0.4\linewidth,clip=}
\end{tabular}
\caption{(Color online) (a) The reflected fraction of the total energy flux and (b) the scattering lengths normalized by the scale $b_0$ as functions of the nondimensional impact parameter $z=b/b_0$ for temperature $T=0.25T_c$. Curves, from top to bottom, correspond to the particle (`number') flux, energy flux, and the net energy flux resulting from temperature gradient [$F$, $F_E$, and $F_T$ in (a), and $K/b_0$, $K_E/b_0$, and $K_T/b_0$ in (b)]. Inset shows the behavior of scattering length for $0\leq z\leq5$.}
\label{fig:3}
\end{figure*}

The reflected fractions of the ``number'' and energy fluxes and the corresponding scattering lengths normalized by $b_0$ are shown, for temperature $T=0.25T_c$, in Fig.~\ref{fig:3} as functions of
$z=b/b_0$. In particular, Fig.~\ref{fig:3}(a) shows the $z$ dependence of the reflected fraction of the total fluxes for relatively small ($0\leq z\leq10$) maximum values of the nondimensional impact parameter. Figure~\ref{fig:3}(b) illustrates the $z$ dependence of the scattering lengths, normalized by $b_0$, for a wide range of impact parameters, with the inset showing $K/b_0$ vs $z$ for relatively small values of $z$ (from 0 to 5). As can be seen from Fig.~\ref{fig:3}(b), for large values of impact parameters (of the order of $100b_0$) the difference between the three scattering lengths becomes noticeable: thus, for $b=200b_0$ and $T=0.25T_c$, the scattering lengths $K_E$ and $K_T$, defined for the energy flux and for the net energy flux resulting from the temperature gradient, are about 7\% and 15\%, respectively, lower than the scattering length $K$ calculated for the number flux. This contrasts our earlier results~\cite{SBBS4} obtained for Andreev cross sections of quantized vortex rings at lower temperatures, $T\leq 150\,\mu{\rm K}$.

\begin{figure*}[ht]
\centering
\begin{tabular}{cc}
\epsfig{file=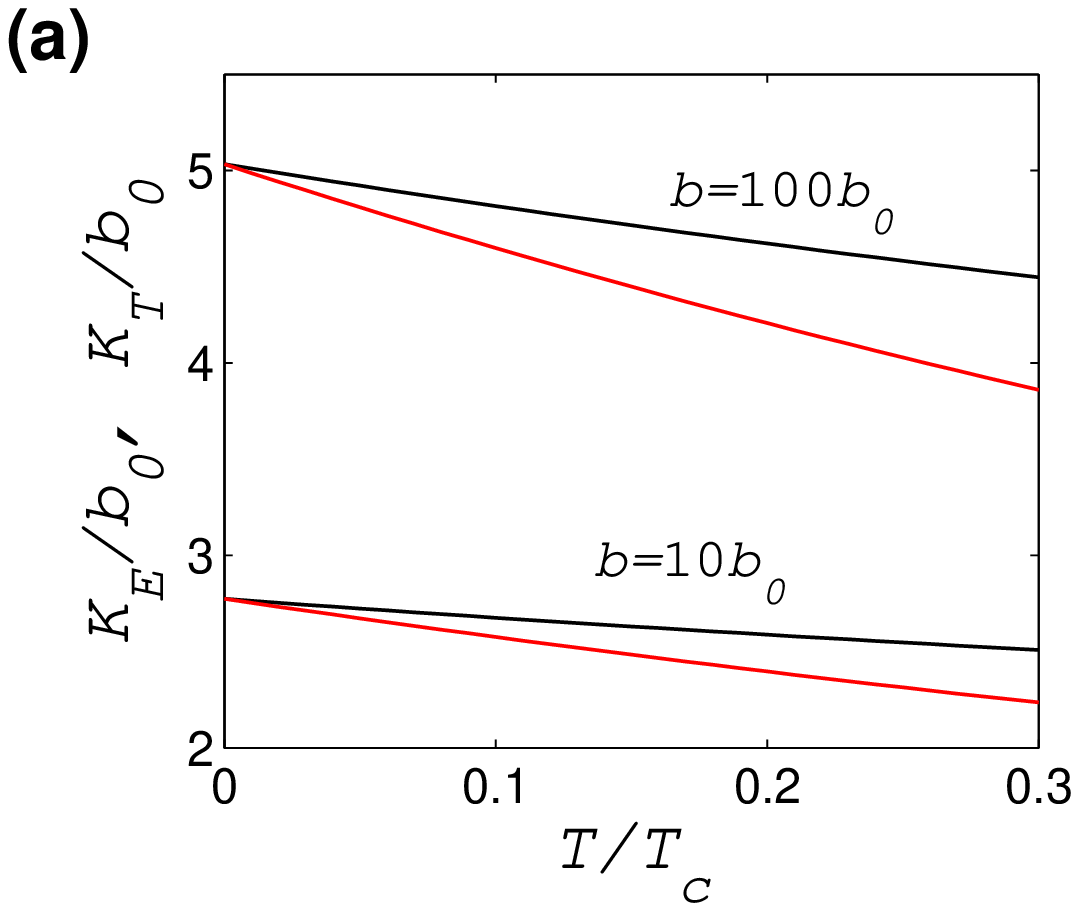,width=0.37\linewidth,clip=} &
\epsfig{file=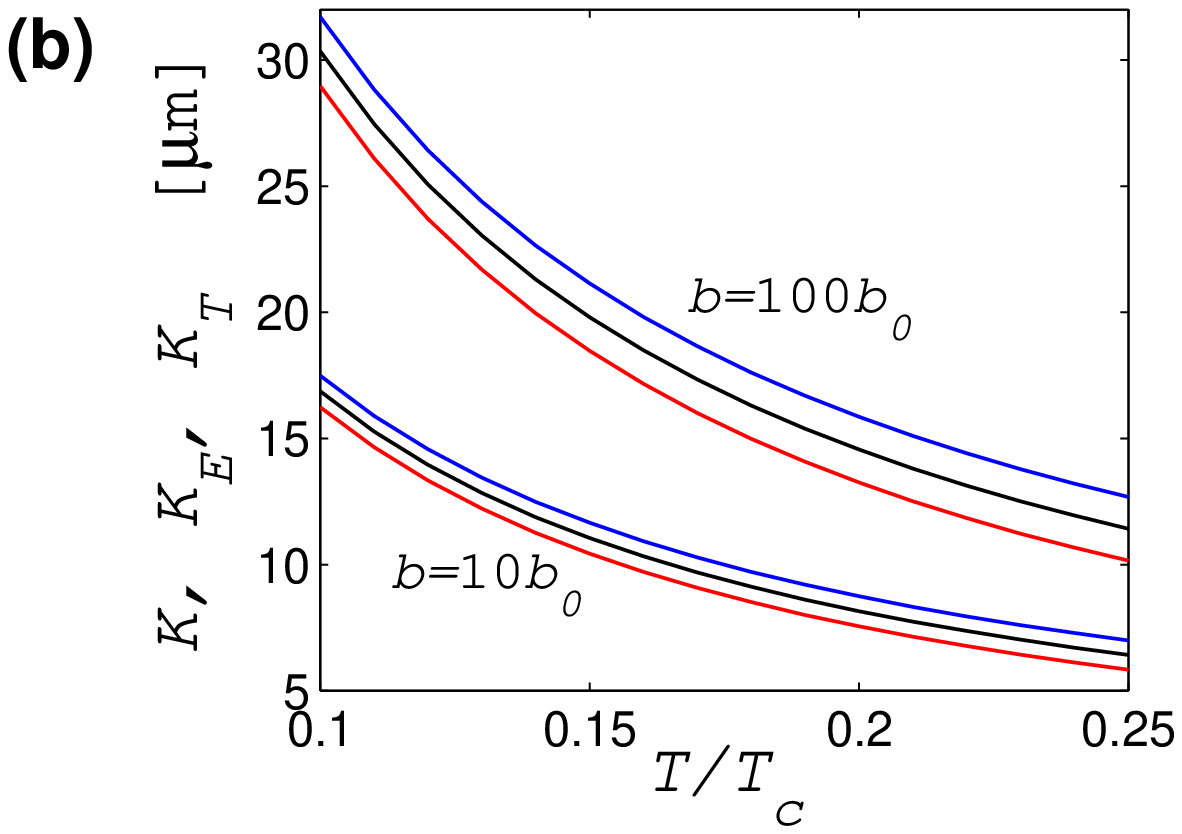,width=0.4\linewidth,clip=}
\end{tabular}
\caption{(Color online) (a) Temperature dependence of the nondimensional scattering lengths, $K_E/b_0$ and $K_T/b_0$ for two values of the impact parameter. In each pair of curves, the upper (black) and the lower (red) lines correspond to $K_E$ and $K_T$, respectively. (b)~Temperature dependence of the dimensional scattering lengths $K$, $K_E$, and $K_T$ [$\mu$m] for 0 bar pressure. In each group of three curves, the top (blue), middle (black), and bottom (red) lines show the behavior of $K$, $K_E$, and $K_T$, respectively.}
\label{fig:4}
\end{figure*}

To better illustrate the temperature dependence of the scattering lengths $K_E$ and $K_T$, Fig.~\ref{fig:4}(a) shows these scattering
lengths, normalized by $b_0$, as functions of temperature for two maximum values of the impact parameter, $b=10b_0$ (the lower pair of curves) and $b=100b_0$ (the upper pair). Figure~\ref{fig:4}(b) illustrates the temperature dependence of the dimensional scattering lengths $K$, $K_E$, and $K_T$ in units of $\mu$m for the same two values of $b$. Note that, for any fixed $b$, the scattering length $K$ for the number flux of excitations behaves with temperature as $1/T$, as follows from Eqs.~(\ref{eq:b0}) and (\ref{eq:Ktilde}).

\section{Conclusions}
\label{sec:conclusions}

The results of this work are applicable in the case where thermal quasiparticle excitations in $^3$He-$B$ propagate ballistically (i.e., when, at low pressure, the temperature is in the range $0\leq T\lesssim0.3T_c$).

For a particle (`number') flux of thermal quasiparticle excitations incident on the rectilinear quantized vortex and propagating in the direction orthogonal to the vortex filament, we calculated exactly the Andreev-reflected fraction $F(b)$ of the total flux of excitations whose impact parameters do not exceed an arbitrary maximum value $b$. We then defined, as $K(b)=2bF(b)$, the scattering length of Andreev reflection. We found that both the fraction $F(b)$ and the scattering length $K(b)$ normalized by the scale $b_0$ of Andreev scattering [see Eq.~(\ref{eq:b0})] are universal functions of the nondimensional impact parameter, $z=b/b_0$. We found that, for $b\gg b_0$, the fraction $F(b)$ decreases with $b$ rather slowly, as $b^{-1}\ln b$, while the scattering length increases with $b$ logarithmically.

We also defined and calculated exactly the scattering lengths: $K_E$ for the flux of energy carried by thermal quasiparticle excitations, and $K_T$ for the net energy flux resulting from the (small) temperature gradient. We analyzed the behavior with the temperature of the scattering lengths $K_E$ and $K_T$ within the temperature range of the ballistic regime. In the low temperature limit (e.g., for the lowest accessible temperature $T=0.1T_c\approx100\,\mu{\rm m}$) the difference between all three scattering lengths $K$, $K_E$, and $K_T$ is small, in agreement with findings of our earlier paper~\cite{SBBS4}. However, this difference becomes more pronounced with increasing of both temperature and the impact parameter $b$: For example, at $T=0.25T_c$ and $b=100b_0$ the scattering lengths $K_E$ and $K_T$ are, respectively, about 7\% and 15\% lower than the scattering length, $K$ defined for the number flux.

It can be expected that, as $T/T_c$ increases, a difference between the scattering length defined for the particle (`number') flux and that defined for the energy flux becomes sufficiently large to be detected experimentally. We note here that it would be experimentally easier and more practical to detect and measure a difference between scattering cross-sections of Andreev reflection from the vortex tangle rather than a difference between scattering lengths of reflection from an individual vortex. Based on the technique~\cite{Fisher2001} developed by the Ultra Low Temperature Group at Lancaster, the experiment can be designed to probe the same vortex tangle by ambient thermal excitations as well as by a purposely generated thermal quasiparticle beam. A nearly homogeneous vortex tangle should be generated between two black box radiators (bolometers). One of the radiators will produce a quasiparticle beam propagating through the tangle, while the other will measure a transmitted energy flux. Several vibrating wire detectors should be located inside and around the turbulent tangle to measure a reflected fraction of the particle (`number') flux, both in the presence of the incident beam of excitations, produced by one of the black box radiators, and in the case where only ambient excitations are present. Such measurements should be carried out for various temperatures of the quasiparticle beam and of the bulk superfluid to enable the experimentalist to compare the properties of the reflected particle and energy fluxes and hence to determine experimentally a difference between the scattering cross sections for these fluxes.

\section*{Acknowledgments}
We thank D. I. Bradley and G. R. Pickett for useful discussions. This research was supported by the Leverhulme Trust, the EPSRC UK, and the European FP7 Programme MICROKELVIN Project no.~228464.

\end{document}